\documentclass[prb,aps,twocolumn]{revtex4}
\usepackage{amsfonts}
\usepackage{amssymb,latexsym}
\usepackage{epsfig}
\usepackage{graphics}

\begin{document}

\title{Quantum versus thermal fluctuations in the harmonic chain and 
experimental implications}
\author {K. Sch\"onhammer}
\affiliation{Institut f\"ur Theoretische Physik, Universit\"at
  G\"ottingen, Friedrich-Hund-Platz 1, D-37077 G\"ottingen}

\date{\today}

\begin{abstract}

The nonzero ground-state energy of the quantum mechanical harmonic oscillator  
implies quantum fluctuations around the minimum of the potential with
the mean square value proportional to  Planck's constant.
 In classical mechanics thermal fluctuations
occur when the oscillator is coupled to a heat bath of temperature $T$.
At finite temperature quantum statistical mechanics allows the description
of the transition from pure quantum fluctuations at $T=0$ to classical
thermal fluctuations in the high temperature limit.
 It was early
pointed out by Peierls that the mean square thermal fluctuations
in a {\it harmonic chain} increase {\it linearly} with the distance of the atoms
in the chain, destroying long range crystalline order. 
The corresponding pure quantum fluctuations  lead to a much
slower {\it logarithmic} increase with the distance from the fixed end
of the chain. 
It is also shown that this implies, fo example, the
absence of sharp Bragg peaks in x-ray scattering in an infinite chain
at zero temperature, which instead show power law behaviour typical
for one dimensional quantum liquids (called {\it Luttinger liquids}).

\end{abstract}
\maketitle

\section{Introduction}

This paper addresses the question
``what are quantum fluctuations,
how do they differ from classical thermal fluctuations
 and what are measurable consequences?''.
In order to simplify the discussion a single particle in a
one dimensional external potential is treated before switching
 to one of the simplest many-body systems, the harmonic chain.

\noindent
Despite the fact that quantum mechanics and quantum statistical
mechanics are used in this paper we start with a short discussion
 in the framework
of classical physics. In classical mechanics
a  particle of mass $m$ moving in a time independent external 
potential $V(x)$ has a well defined ``ground state'' if the potential
is everywhere larger than its value at a single (non degenerate)
mimimum. The typical example is the harmonic oscillator
\begin{equation}
\label{hopotential}
V(x)=\frac{\lambda}{2}x^2=\frac{1}{2}m\omega_0^2x^2
\end{equation}
where the particle at rest at the origin corresponds to the
ground state. Thermal fluctuations around this position occur
if the particle is coupled to a heat bath described by  
a canonical ensemble. Then the probability distribution to 
find the particle at position $x$ is given by
$w_{T,cl}(x)\sim \exp(-\beta V(x))$, where $\beta=1/(k_BT)$ with
$k_B$ the Boltzmann constant and $T$ the temperature of
the bath \cite{Huang}.  
The moments $\langle x^n\rangle_{T,cl}\equiv \int x^nw_{T,cl}(x)dx$
for $n=1,2$ describe the fluctuations around the minimum.
As $w$ is an even function $\langle x\rangle_{T,cl}$ vanishes 
and $\langle x^2\rangle_{T,cl}$ directly gives the mean square deviation.
In order to obtain $\langle x^2\rangle_{T,cl}$
one can either use the fact that a Gaussian
probability distribution has the form 
$w_G(x)\sim \exp(-x^2/(2\langle x^2\rangle))$
or one can use
the equipartition theorem 
of classical statistical mechanics \cite{Huang}
 $\langle V\rangle_{T,cl}=k_BT/2$
which holds for the quadratic oscillator potential, i.e.
 \begin{equation}
\label{equipart}
\langle x^2\rangle_{T,cl}=\frac{k_BT}{\lambda}=
\frac{k_BT}{m\omega^2_0}\equiv \sigma^2_{cl}~.
\end{equation}
Now we switch to quantum mechanics. The Hamiltonian $\hat H$
usually depends on operators $\hat A,\hat B,...$ which do {\it not}
commute. A simple example is a particle in an external potential 
\begin{equation}
\hat H=\frac{\hat p^2}{2m}+V(\hat x)~,
\end{equation}
where the commutation relation $[\hat x,\hat p]=i\hbar \hat1$
leads to the uncertainty relation \cite{Baym}
 \begin{equation}
\label{uncertainty}
\Delta\hat x\Delta \hat p\ge \hbar/2~.
\end{equation}
Here $\Delta \hat A\equiv \sqrt{\langle \hat A^2\rangle_\Phi
-\langle \hat A\rangle^2_\Phi}$ with $ \langle \hat A^n\rangle_\Phi
\equiv \langle \Phi|\hat A^n|\Phi \rangle$ and $|\Phi \rangle$
is the quantum state of the system. 

For the harmonic oscillator
the uncertainty relation implies quantum fluctuations even in the
ground state $|0\rangle$,
 usually called ``zero point motion''. The corresponding
probability distribution $w_0(x)$ is again Gaussian with vanishing
$\langle \hat x\rangle_0$ (see next section). The mean square
fluctuations can be obtained without the explicit form of ground state
wavefunction using the {\it virial theorem} \cite{Merzb}. 
For the eigenstates
of the harmonic oscillator it states the equality of the expectation
value of the kinetic energy and the potential energy. 
With the ground state energy $\epsilon_0=\hbar\omega_0/2$ this implies
 \begin{equation} 
\label{ground statefl} 
\langle \hat x^2\rangle_0=\frac{\hbar }{2m\omega_0}\equiv \sigma^2_{qm}~.
\end{equation}
In section II we dicuss how the results for the classical
thermal fluctuation Eq. (\ref{equipart}) and the quantum fluctuation
Eq. (\ref{ground statefl}) connect as a function
of temperature. \cite{Messiah}.  This discussion is
extended to the harmonic chain in section III. 
In the classical ground state of the chain the $n$-th atom is located at 
the lattice position $na$ if atom number zero is fixed at the origin.
Here $a$ is the lattice constant. It was pointed out early by Peierls  
\cite{Peierls} in the classical context that while $\langle
x_n\rangle_{cl}=na$ holds also at finite temperatures, thermal
fluctuations destroy long range order at any finite temperature as
the mean square deviations $\langle(x_{n+l}-x_l-na)^2\rangle$
of the separations of two atoms diverge linearly with $n$ and
proportional to the temperature $T$. Therefore crystalline order
exists classically only at zero temperature. It is discussed in
section III how quantum fluctuations destroy crystalline order
even at $T=0$ by a much weaker logarithmic divergence with $n$.
It is again discussed how the results for the results for the
classical thermal fluctuations and the pure quantum fluctuations
connect as a function of temperature. 
It turns out that
the different dependence on the frequency in
Eqs. (\ref{equipart}) and  (\ref{ground statefl}) plays a decisive role. 

In the context of harmonic lattices in {\it two} dimensions, a similar
logarithmic divergence occurs in the calculation of classical thermal
fluctuations.\cite{Jancovici,MS}
The main results presented here for the harmonic chain,
including the power law shape of the Bragg peaks, cannot be found
in the literature.

\section{The harmonic oscillator}

As a warm-up to analyzing the harmonic chain, we first discuss both the
quantum fluctuations and the thermal fluctuations of a single harmonic
oscillator. Our treatment uses the ladder operators that are introduced
in almost every textbook. \cite{Baym,Merzb,Messiah}

\subsection{Ground state properties}

\noindent The Hamiltonian of a one-dimensional harmonic oscillator reads
\begin{equation}
\hat H =\frac{1}{2m}\hat p^2+\frac{\lambda}{2}\hat x^2 ,
\end{equation}
where $m$ is the mass of the particle and $\lambda$ the spring
constant. With the frequency $\omega_0=\sqrt{\lambda/m}$ one defines 
the lowering operator $\hat a$ and its adjoint $\hat a^\dagger$
\begin{equation}
\label{defa} 
\hat a =\sqrt{\frac{m\omega_0}{2\hbar}}\hat x +
\frac{i}{\sqrt{2m\hbar \omega_0}}\hat p~;~~ \hat a^\dagger =
\sqrt{\frac{m\omega_0}{2\hbar}}\hat x -
\frac{i}{\sqrt{2m\hbar \omega_0}}\hat p~,
\end{equation}
which obey the commutation relation $[\hat a, \hat a^\dagger ]=\hat 1$.
 The position operator $\hat x$ and the 
momentum operator $\hat p$ read in terms of $a$ and $\hat a^\dagger$
\begin{equation}
\label{xp}
\hat x =\sqrt{\frac{\hbar}{2m\omega_0}}\left (\hat a+ \hat a^\dagger
\right);~~\hat p=-i\sqrt{\frac{m\hbar \omega_0}{2}}\left(\hat a-
 \hat a^\dagger \right ).
\end{equation}
The Hamiltonian $\hat H$, then takes the form
\begin{equation}
\label{all} 
\hat H=\hbar \omega_0 \left( \hat a^\dagger \hat a+\frac{1}{2}\right )
\end{equation}
and  its eigenstates $|n\rangle$ and eigenvalues 
 are given by
\begin{equation}
\label{evev}
|n\rangle =\frac{(\hat a^\dagger)^n}{\sqrt{n!}}|0\rangle~;~~
\epsilon_n=\hbar \omega_0 \left(n+\frac{1}{2} \right ).
\end{equation}
 The ground state $|0\rangle$  is annihilated by $\hat a$, i.e.
  $\hat a|0\rangle
=0$ holds. In the position representation $\langle x|\hat a|0\rangle=0$
is a linear differential equation which determines the
ground state wavefunction\cite{Baym}
\begin{equation}
\label{phi0}
\phi_0(x)\equiv \langle x|0\rangle
=\left( \frac{m\omega_0}{\pi \hbar}\right)^{1/4}
\exp{\left (-\frac{m\omega_0}{2\hbar}x^2\right )}~.
\end{equation}
In order to calculate ground state expectation values of functions of $\hat x$
and $\hat p$ one can either
use the explicit form of $ \phi_0(x)$
or the property    $\hat a|0\rangle=0$ only.
As an example we consider the operator $\exp{(-ik\hat x)}$.
Its expectation value is readily calculated in the position representation. 
It is given by the Fourier transform
of $|\phi_0(x)|^2$ which is obtained by a Gaussian integration.
 Alternatively one can use
Eq. (\ref{xp}) and the {\it Baker-Haussdorff(BH) formula} \cite {Merzb} 
which will be used also for the harmonic chain. This formula reads
\begin{equation}
e^{\hat A +\hat B} =e^{\hat A}e^{\hat B}e^{-\frac{1}{2}[\hat A,\hat B]},~~\mbox{if}~~
[\hat A,[\hat A,\hat B]]=0=[\hat B,[\hat A,\hat B]].
\end{equation}
For operators $\hat A$ and $\hat B$ {\it linear} in the ladder
operators the requirements are fulfilled. 
 With $v\equiv -ik\sqrt{\hbar/(2m\omega_0)}$ one obtains
\begin{eqnarray}
\langle 0|e^{-ik\hat x}|0\rangle &=& \langle 0|e^{v\hat
  a^\dagger+v\hat a}|0\rangle\nonumber \\
 &=& \langle 0|e^{v\hat a^\dagger}e^{v\hat a}|0\rangle
e^{-\frac{1}{2}v^2[\hat a^\dagger,\hat a]}\nonumber \\
 &=& \exp{\left (-\frac{\hbar k^2}{4m\omega_0}\right )}~.
\label{eikx}
\end{eqnarray}
Because of $\hat a|0\rangle=0$ which implies
  $\langle 0|\hat a^\dagger=0 $ the expectation value in the second equality
equals $1$. As a test we recover the ground state density $w_0(x)$
which is obtained much more simply by squaring $\phi_0(x)$.
The operator $\hat \rho(x)$ of the particle density can be 
written as an operator valued Dirac delta function
 \begin{eqnarray}
\hat \rho(x)&=&|x\rangle\langle x|=\int dx'|x'\rangle\langle
x'|\delta(x-x') \nonumber \\
&\equiv& \delta(x-\hat x)~. 
\end{eqnarray}
 Using the representation
of the Dirac delta function as a Fourier integral one has to perform
 a Gaussian integration
\begin{eqnarray}
\label{Gauss}
w_0(x) &=&\langle 0|\delta(x-\hat x)|0\rangle=\frac{1}{2\pi}
\int_{-\infty}^\infty e^{ikx}
\langle 0|e^{-ik\hat x}|0\rangle dk \nonumber \\
 &=&\left( \frac{m\omega_0}{\pi \hbar}\right)^{1/2}
\exp{\left (-\frac{m\omega_0}{\hbar}x^2 \right )}~.
\label{density0}
\end{eqnarray}
For finite temperatures the corresponding calculation is {\it simpler} than via
 the position representation.

Because of $[\hat a,\hat H]=\omega_0\hat a$ the time dependence of 
the lowering operator $\hat a$ in the Heisenberg picture takes the simple form
\cite{Baym}
\begin{equation}
\label{timedep} 
\hat a(t)=e^{i\hat Ht/\hbar}\hat ae^{-i\hat Ht/\hbar}=\hat ae^{-i\omega_0t}~.
\end{equation}

\subsection{Finite temperature properties}

We consider the harmonic oscillator in thermal equilibrium
 described by the canonical ensemble
with temperature $T$. As discussed in textbooks on statistical 
mechanics \cite{Huang} the expectation value
of an observable $A$ in thermal equilibrium is given by
\begin{equation}
\label{canens}
\langle \hat A \rangle =\frac{\mbox{Tr}( \hat Ae^{-\beta \hat H}) }
{\mbox{Tr}(e^{-\beta \hat H})}
=\frac{1}{Z}\sum_n \langle E_n|\hat A|E_n \rangle e^{-\beta E_n},
\end{equation}
where $\beta=1/(k_BT)$, the 
$|E_n\rangle $ are the eigenstates
of the Hamiltonian and $Z= \mbox{Tr}(e^{-\beta \hat H})=  \sum_n
e^{-\beta E_n}$ is
 the partition function.
 
It is useful to consider expectation values of  
an important class of operators,   $ (\hat a^\dagger)^l\hat a^n$,
where $l$ and $n$ are integers. 
As only diagonal 
matrix elements contribute in Eq. (\ref{canens}) the thermal
expectation values $\langle (\hat a^\dagger)^l\hat a^n
\rangle$ vanish unless $l=n$.
 We therefore calculate 
the expectation values of the operators 
$ \hat A_n \equiv  (\hat a^\dagger)^n\hat a^n$
\begin{eqnarray}
\langle \hat A_n\rangle &=& \frac{1}{Z}\mbox{Tr}
\left((\hat a^\dagger)^n\hat a^n e^{-\beta \hat H}\right)\nonumber \\
 &=& \frac{1}{Z}\mbox{Tr} \left((\hat a^\dagger)^n\hat a^{n-1} 
e^{-\beta \hat H}e^{\beta \hat H}\hat a
e^{-\beta \hat H}\right)\nonumber \\
 &=& \frac{1}{Z}\mbox{Tr} \left((\hat a^\dagger)^n\hat a^{n-1} 
e^{-\beta \hat H}\hat a e^{-\beta\hbar \omega_0}\right) \nonumber \\
 &=& e^{-\beta \hbar \omega_0} \frac{1}{Z}\mbox{Tr}\left( \hat a
 (\hat a^\dagger)^{n}\hat a^{n-1} e^{-\beta \hat H}\right).
\label{Wick1}
\end{eqnarray}
In the third line the result $ \hat a e^{-\beta \hbar \omega_0}$
for the Heisenberg operator with imaginary argument $\hat a (-i\beta \hbar)$
was inserted and in the last equality the cyclic invariance of the
 trace was used. 
Next, the $\hat a$ on the left of creation operators is moved back to the right.
With the operator identity
\begin{eqnarray}
\hat a (\hat a^\dagger)^n=[ \hat a, (\hat a^\dagger)^n]+ (\hat a^\dagger)^n
\hat a=n (\hat a^\dagger)^{n-1}+(\hat a^\dagger)^n
\hat a
\end{eqnarray}
 valid for $n\ge 1$, Eq. (\ref{Wick1}) goes over to the recursion relation
\begin{eqnarray}
\langle \hat A_n\rangle &=& e^{-\beta \hbar \omega_0}
\left ( n\langle \hat A_{n-1}\rangle+\langle \hat A_n\rangle
\right )\nonumber \\
 &=& \frac{n}{e^{\beta \hbar \omega_0}-1}\langle \hat A_{n-1}\rangle
\end{eqnarray}
for  $n\ge 1$. With the trivial starting point $ \langle \hat A_0\rangle=1  $ 
one obtains for $n=1$ the well known result
\begin{equation}
\langle \hat a^\dagger \hat a \rangle = \frac{1}{e^{\beta \hbar
    \omega_0}-1}
\equiv n_B(\omega_0),
\end{equation}
where $n_B$ is the Bose function. This result can be obtained more
directly using the thermodynamic relation $\langle \hat H\rangle=
-\partial \log{Z}/\partial \beta$.
 For general $n>1$
one obtains {\it Wick's theorem} \cite{FW} for the harmonic oscillator
\begin{equation}
\label{Wicktheorem}
\langle (\hat a^\dagger)^n\hat a^m \rangle =\delta_{nm} n!
\langle \hat a^\dagger \hat a \rangle ^n.
\end{equation}
Generalizations of Wick's theorem 
play an important role in quantum field theory.

Wick's theorem, Eq. (\ref{Wicktheorem}), can be used to obtain the
relation
\begin{eqnarray}
\langle e^{\lambda \hat a ^\dagger}e^{\mu \hat a} \rangle
&=&\sum_{n=0}^\infty \sum_{m=0}^\infty
\frac{\lambda^n\mu^m}{n!m!}\langle
 (\hat a^\dagger)^n\hat a^m \rangle
 =\sum_{n=0}^\infty\frac{(\lambda\mu)^n}{n!}\langle 
\hat a^\dagger \hat a \rangle^n \nonumber \\
&=&e^{\lambda\mu \langle \hat a^\dagger \hat a \rangle}=e^{\lambda\mu n_B(\omega_0)}~.
\label{no2}
\end{eqnarray}
which is used frequently in the following.
Operator products like those on the lhs of Eqs. (\ref{Wicktheorem}) and
 (\ref{no2})
are called ``normal ordered'' because all lowering operators, in
the field theoretical context
called {\it annihilation} operators, are to the right of the
 {\it creation} operators $\hat a^\dagger$. 
Products like $(\lambda \hat a^\dagger +\mu \hat a)^2$
are {\it not} normal ordered. 
In the calculation of expectation values of operators of this type
one uses the commutation relation $[\hat a,\hat a^\dagger]=\hat 1$
 for normal ordering
\begin{eqnarray}
\label{bilinear}
\langle (\lambda \hat a ^\dagger +\mu \hat a)^2\rangle
&=& \langle\lambda^2(\hat a^\dagger)^2+
\lambda\mu(\hat a^\dagger\hat a+\hat a\hat a^\dagger)+\mu^2\hat a^2
\rangle  \nonumber \\
&=& \lambda\mu(1+2n_B) ,
\end{eqnarray}
which yields the important identity\cite{Mermin}
\begin{equation}
\label{nonno}
\langle e^{\lambda \hat a ^\dagger+\mu \hat a} \rangle=
\langle e^{\lambda \hat a ^\dagger}e^{\mu \hat a} \rangle 
e^{-[\lambda \hat a ^\dagger,\mu \hat a]/2}
=e^{\langle (\lambda \hat a ^\dagger +\mu \hat a)^2\rangle/2},
\end{equation}
where we have used the BH-identity as well as Eqs. (\ref{no2}) and 
(\ref{bilinear}).

\noindent  With this identity
one can immediately generalize the calculation of the average density
 in Eq. (\ref{density0})
using $\langle \exp{(ik\hat x)}\rangle=\exp{(-k^2\langle \hat x^2
  \rangle /2)}$.
 This implies
that the average density is Gaussian for all temperatures
\begin{equation}
\langle \delta (x-\hat x)\rangle =\frac{1}{\sqrt{2\pi\langle \hat x^2 \rangle }}
\exp{\left (-\frac{x^2}{2\langle \hat x^2 \rangle}\right )},
\end{equation}
with 
\begin{equation}
\label{x2mittel}
\langle \hat x^2 \rangle=\frac{\hbar}{2m\omega_0}
\left [1+2n_B(\omega_0) \right].
\end{equation}
This derivation of the result for the average density at arbitrary
temperature is much shorter than the one presented in Ref. 4.
 At $T=0$ the Bose function $n_B(\omega_0)$ vanishes
 and one obtains 
the ground state result Eq.(\ref{ground statefl}).
In the high temperature limit $\epsilon\equiv \hbar \omega_0/k_BT\ll 1$ one
obtains $1+2n_B(\omega_0)=1+2(\epsilon+\epsilon^2/2+...)^{-1}
=2k_BT/\hbar\omega_0+O(\epsilon)$ and $ \langle \hat x^2 \rangle  $
goes over to the classical result Eq. (\ref{equipart}). The crossover
from the pure quantum fluctuations at $T=0$ to the classical thermal
fluctuations is shown in Fig. 1. At any nonzero temperature quantum
and thermal fluctuations cannot be disentangled, 
but for $ k_BT/\hbar \omega_0 \ll 1$ the fluctuations are ``quantum
dominated'' and for  $k_BT/\hbar \omega_0 \gg 1$ they are almost like
classical thermal fluctuations.

 The generalization 
of Eq. (\ref{x2mittel}) to the case of the harmonic chain 
plays an important role in the next section.

\begin{figure}
\label{figho}
\centering
\epsfig{file=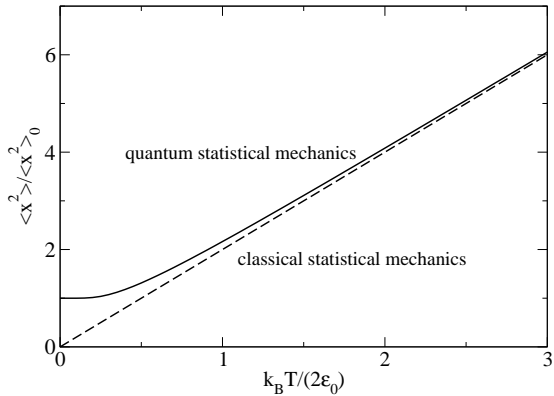,angle=-90,scale=.3}
\vspace{0.2cm}
\caption { Mean square deviation $\langle \hat x^2 \rangle$ scaled by
 the ground state value of the
  harmonic oscillator as function of the
scaled temperature $k_BT/(\hbar \omega_0)$.}
\end{figure}
 Using the BH-identity the relation 
$ \langle e^{\hat A}\rangle =e^{\langle \hat
  A^2\rangle/2}$ 
of Eq. (\ref{nonno}) is easily generalized to 
\begin{equation}
\label{identities}
\langle e^{\hat A}\ e^{\hat B} \rangle
=e^{(\langle \hat A^2+ \hat B^2+2\hat A \hat B \rangle )/2}
\end{equation}
for any operators $\hat A$ and  $\hat B$
that are {\it linear} in the ladder operators.

\section{The harmonic chain}

We now apply the same principles to a {\it harmonic chain}: a
one-dimensional chain of $N$ coupled harmonic oscillators (e.g.
masses connected by springs).
This system is often discussed in classical
mechanics courses, because it allows a complete
 analytical solution.
In solid state physics books the harmonic chain usually serves as an
 introduction to chapters on
lattice dynamics of solids \cite{AM}.
  The interaction between atoms consists of a short
range repulsive and a long range attractive interaction resulting in a
two-body
interaction potential with a deep minimum.
Therefore the atoms are assumed to be
confined to their wells and are treated as {\it distinguishable},
 i.e. the fact 
whether the atoms are fermions or bosons does not enter the description.
 In the harmonic approximation
the atoms are assumed to perform small oscillations around these
 minima which leads
to {\it linear} equations of motion. For low temperatures $T$ quantum
effects are important
resulting in the Debye law  $\sim T^d$ for the specific heat,
 where $d$ is the spatial
dimension of the lattice \cite{AM}. It was pointed out early by 
Peierls in the classical context that
 fluctuations have drastic effects for $d\le 2$, like the loss of long
 range order  \cite{Peierls,Jancovici,MS}.
Here we discuss the harmonic chain ($d=1$) in the framework 
of quantum statistical mechanics.

\noindent An exact analytical calculation of correlation functions
like the  {\it static structure factor} \cite{AM} which appears
in the theoretical description of  X-ray scattering is
 possible for the harmonic chain
 using the operator identities
introduced in the context of the harmonic oscillator.
 Power law behaviour emerges in limiting regions.  
Similar power laws show up for interacting fermions
 in one dimension. This is called ``Luttinger liquid'' behaviour
\cite{Tomonaga,Luttinger,Haldane,SM}.

\subsection{Normal modes and ladder operators}

\noindent In most textbooks chains with periodic boundary conditions
are discussed corresponding  to particles on a ring \cite{AM}.
 For the discussion of
the quantum fluctuations it is more convenient to  
 ``pin'' one end of the chain (``sample holder'').
 For a chain of equal masses $m$ and spring constants $\lambda$
  the
Hamiltonian $\hat H=\hat T+\hat V$ reads 
\begin{eqnarray}
\label{Hamiltonian}
\hat H  =\sum_{i=1}^N\frac{\hat p_i^2}{2m}+
\frac{\lambda}{2}\left [(\hat x_1-a)^2+
\sum_{i=1}^{N-1}(\hat x_{i+1}-\hat x_i-a)^2\right]
\end{eqnarray}
 when the left end of the
chain is pinned to the origin by a harmonic force of the same strength
as the interparticle force .
Here $a$ is the separation at which the two-body interaction has its
minimum. The potential $V$ vanishes when the $\hat x_n$ are replaced
by the classical ground state positions $na$. 
If one introduces the displacement operators
\begin{equation} 
 \hat u_n\equiv \hat x_n-na
\end{equation} 
 the operator of the potential energy reads
\begin{equation} 
\label{pot}
  \hat V =\frac{\lambda}{2}      \left [ 
\hat u_1^2+ \sum_{i=1}^{N-1}(\hat u_{i+1}-\hat u_{i})^2\right ]
\equiv \frac{\lambda}{2}\sum_{n,m} C_{nm}\hat u_n \hat u_m~,
\end{equation}
where the last equality defines the matrix $ {\bf C} $.
In order to obtain the normal
modes for this equal mass chain one solves the eigenvalue problem
  \begin{equation}
 {\bf C} {\bf \xi}^{(j)}=C_j{\bf \xi}^{(j)}~.
\label{eigen}
\end{equation}  
 and expands the displacement operators $\hat u_n$ into normal
 modes $\hat U_j$
as in the classical case
\begin{equation}
\hat u_n=\sum_{j=1}^N 
\hat U_j\xi_n^{(j)}~;~~~~~\hat U_j=\sum_{n=1}^N \xi_n^{(j)}\hat u_n~.
\label{nomo}
\end{equation}   
The components $ \xi_n^{(j)}$ of the
 orthonormal eigenvectors $ {\bf \xi}^{(j)} $
of the real symmetric eigenvalue problem Eq. (\ref{eigen}) are chosen real.
   With the corresponding normal 
momentum operators  $\hat P_j=\sum_{n=1}^N \xi_n^{(j)}\hat p_n$
 the Hamiltonian takes the form
\begin{equation}
\label{HPU}
\hat H=\sum_{j=1}^N \left[\frac{\hat P_j}{2m}+\frac{\lambda}{2}
  C_j\hat U_j^2\right ]
 \end{equation}
and $[\hat U_i,\hat P_j]=i\hbar \delta_{ij}$ holds. In the normal mode
basis the system looks like a system of {\it independent}
harmonic oscillators with
eigenfrequencies given by
\begin{equation}
\omega_j^2=\frac{\lambda}{m}C_j\equiv \omega^2_sC_j~,
\end{equation}
where the index ``s'' indicates the case of a single oscillator (in
the previous section $\omega_s$ was labeled $\omega_0$).
 
The explicit form of the matrix ${\bf C}$ 
can be read off Eq. (\ref{pot}). The
equations for the components $ \xi_n^{(j)}$ of the eigenvector
 ${\bf \xi}^{(j)}$  read for $2\le n\le N-1$
\begin{equation}
\label{EV}
2\xi_n^{(j)}-\xi_{n-1}^{(j)}-\xi_{n+1}^{(j)}
=C_j\xi_n^{(j)}
\end{equation}
and the two boundary equations are given by
\begin{equation}
\label{EVRand}
2\xi_1^{(j)}-\xi_2^{(j)} 
=C_j\xi_1^{(j)},~~~
\xi_N^{(j)}-\xi_{N-1}^{(j)} 
=C_j\xi_N^{(j)}~.
\end{equation}
These equations can be solved with the ansatz
\begin{equation}
\xi_n^{(j)} =\xi_n(\tilde k_j) \sim \mbox{Re}[e^{i(\tilde k_jn+\phi_j)}],
\end{equation}
where the $\tilde k_j$ and $\phi_j$
  are dimensionless real numbers and ``$\mbox{Re}$''
denotes the real part.  We formally use this 
ansatz also for $n=0$ and $n=N+1$. 
It is shown below that conditions on the  $ \xi_n(\tilde
k_j)$ at the boundaries can be formulated
such that the two boundary equations take
the same form as 
Eq. (\ref{EV}). Therefore  
the eigenvalues of ${\bf C}$ follow  from the ansatz as
\begin{equation}
 C_j=2(1-\cos\tilde k_j)=  4\sin^2(\tilde k_j/2)
\label{Ansatz}
\end{equation}
and the $N$ allowed values for $\tilde k_j$ are determined by the
conditions from the boundary
equations. 

\noindent The boundary equation Eq. (\ref{EVRand})
at the left end of the chain takes
 the ``bulk form'' Eq. (\ref{EV})
 if one subtracts $ \xi_0(\tilde k_j)$ on the lhs  imposing  
\begin{equation}
\xi_0(\tilde k_j)=0~.
\end{equation}
  This condition is usually called 
``fixed boundary condition'' (fbc).
In order for the eigenvector component to obey this condition one has to
to take $\phi_j=\pi/2$ in the ansatz, i.e.
\begin{equation}
 \xi_n(\tilde k_j) \sim \sin(\tilde k_jn)~.
\end{equation}
 The allowed
$\tilde k_j$-values follow from the boundary equation 
 at the right end of the chain Eq. (\ref{EVRand}) .
 It takes the bulk form Eq. (\ref{EV}) by adding
  $ \xi_N(\tilde k_j) -\xi_{N+1}(\tilde k_j)$ on the lhs
imposing
\begin{equation}
 \xi_N(\tilde k_j)=\xi_{N+1}(\tilde k_j).
\end{equation}
 This is called ``open boundary condition'' (obc).
 Using
$\sin[ \tilde k_j(N+1)]-\sin(\tilde k_jN)=2\cos[\tilde k_j(N+1/2)]
\sin(\tilde k_j/2)$
 one obtains
the allowed values of $\tilde k_j$ and the corresponding (squared)
eigenfrequencies as
\begin{equation}
\label{EWexplicitly}
 \tilde k_j=\frac{j-1/2}{N+1/2}~\pi~;~~~~~~
\omega^2_j=4\omega^2_s\sin^2(\tilde k_j/2)~,
\end{equation}
where $j=1,...,N$.
 The corresponding normalized eigenvectors are given by
 \begin{equation}
\label{EVexplicitly}
\xi_n^{(j)}=\sqrt{\frac{2}{N+1/2}} ~\sin(\tilde k_jn)
\equiv A_N \sin(\tilde k_jn)  ~.
\end{equation}
The normalization constant $A_N$ is obtained using 
$\sum_{n=1}^N\cos(2\tilde k_j n)=Re\left(\sum_{n=1}^N e^{i2\tilde k_j n}\right)$ 
and evaluating the geometric series.

\noindent In the long wave limit $0<\tilde k_j\ll 1$ the
eigenfrequencies $\omega_j=\omega_sC_j $ 
depend {\it linearly} on the $\tilde k_j$
 \begin{equation}
\label{lowenergy}
 \omega_j=2\omega_s\sin(\tilde k_j/2)
\equiv  \frac{2c}{a} \sin(ak_j/2)\approx ck_j.
\end{equation}
Here
$c=a\omega_s$ is the {\it sound velocity}
 and the $k_j\equiv \tilde k_j/a$ are
 the {\it wave numbers}. This ``low energy behaviour'' is
 responsible for the divergences of the correlations 
in the limit $N\to \infty$ studied in the next section.
These divergences do {\it not} occur if each atom moves in an 
{\it additional} harmonic potential of strength $\lambda_{loc}>0$
centered at its lattice position. 
It is a simple exercise to show that the corresponding eigenfrequencies
are given by $(\lambda_{loc}/m+ 4\omega^2_s\sin^2(\tilde k_j/2))^{1/2} $
and stay {\it finite} for $\tilde k_j\to 0$. Except for the
discussion in the outlook we keep $\lambda_{loc}=0$ in the
following calculations.

 As $\omega_j^2 >0$ holds for all eigenfrequencies 
 one can introduce ladder operators as for the case of a single
oscillator
\begin{equation}
\hat a_j=\sqrt{\frac{m\omega_j}{2\hbar}}~\hat
U_j+\frac{i}{\sqrt{2m\hbar
 \omega_j}}\hat P_j
\end{equation}
and the corresponding $\hat a_j^\dagger$ which obey the commutation
relations
\begin{equation}
[\hat a_i,\hat a_j^\dagger]=\delta_{ij}\hat 1~, ~~~[\hat a_i,\hat a_j]=0~.
\end{equation}
 Expressed in terms of these ladder operators the Hamiltonian reads
\begin{equation}
\hat H=\sum_{j=1}^N\left [\hbar \omega_j\left (\hat a_j^\dagger \hat
    a_j+1/2\right )\right ] \equiv \sum_{j=1}^N \hat H_j~.
\label{Hchain}
\end{equation}
Because of the factorization 
$e^{-\beta \hat H}=\prod_{j=1}^N e^{-\beta \hat H_j}$
the operator relations derived for a single harmonic oscillator
can easily be generalized 
to the harmonic chain.

\subsection{Static correlation functions}

 In terms of the ladder operators of the eigenmodes
the operators for the displacements read
\begin{equation}
\hat u_n=\sum_j\sqrt{\frac{\hbar}{2m\omega_j}}
(\hat a_j+\hat a_j^\dagger)\xi_n^{(j)}~.
\end{equation}
The averages $\langle \hat u_n\rangle$ vanish,
 i.e.  the average positions $\langle \hat x_n\rangle=na$
 are given by
the classical $T=0$ result. 
Using $\langle a_j^\dagger\hat a_i\rangle=
\delta_{ij}\langle a_i^\dagger\hat a_i\rangle$ one obtains
the mean square  displacements of the particle
positions as
\begin{eqnarray}
\label{unl}
\langle \hat u_n^2\rangle = \sum_j \frac{\hbar}{2m\omega_j}
 \left [1+2n_B(\omega_j) \right ] (\xi_n^{(j)})^2. 
\end{eqnarray}
As in the introduction we start the discussion
 with the classical limit
$k_BT\gg \hbar \omega_s$ implying
$1+2n_B(\omega_j)\to 2k_BT/\hbar \omega_j$, for {\it all} frequencies 
$\omega_j$, i.e.
\begin{equation}
\label{unlcl}
\langle u_n^2\rangle_{cl} = k_BT\sum_j
\frac{(\xi_n^{(j)})^2}{m\omega_j^2}
=\frac{k_BT}{m\omega_s^2}({\bf C}^{-1})_{nn}~.
\end{equation}
One can either use the explicit form of the eigenfrequencies
Eq. (\ref{EWexplicitly})
 and the eigenvector components Eq. (\ref{EVexplicitly})
 to obtain the mean square fluctuations 
or  calculate
$({\bf C}^{-1})_{nn}$ directly.
The inverse matrix ${\bf C}^{-1}$ can be obtained
 by a simple observation.
The application of ${\bf C}$ to an arbitrary vector ${\bf b}$ can
be read off the lhs of Eqs. (\ref{EV}) and ({\ref{EVRand}).
For ${\bf b}^T_{n}=(1,2,...n-1,n,n,....n)$ one obtains
 ${\bf Cb}_n={\bf e}_{n}$ where ${\bf e}_{n}$ is 
 the $n$-th unit vector. Therefore  ${\bf b}_{n}$ is the  $n$-th 
column of ${\bf C}^{-1}$. The diagonal elements are given
by $({\bf C}^{-1})_{nn}=n$. Using the sound velocity $c=\omega_sa$ this leads to
\begin{equation}
\label{chaincl}
\langle (x_n-na)^2\rangle_{cl}/a^2 =\frac{k_BT}{mc^2}n\equiv \eta_{cl}n
\end{equation}
in the classical limit. Here we compare the fluctuations
to the squared lattice distance and introduced the dimensionless
temperature variable $\eta_{cl}$, which is independent of quantum 
properties. In order for the harmonic approximation to be applicable
$\langle u_1^2\rangle_{cl}$ should be much smaller than $a^2$, i.e.
  $\eta_{cl}\ll 1$ should hold.

\noindent It provides additional insight to consider also the $j$-sum
 on the rhs of 
Eq. ({\ref{unlcl}) which is proportional to $\sum_j\sin^2(\tilde k_j n)/
(4\sin^2(\tilde k_j/2))$. Except for small $n$ the sum is dominated
by the $0<\tilde k_j \ll 1$ contributions. Therefore working
in the numerical evaluation with 
the {\it linearized} dispersion, i.e. $2\sin(\tilde k_j/2) \to \tilde k_j$,
only leads to a small shift compared to the
exact result. This can be seen explicitely in the infinite-chain
limit $N\to \infty$. In this limit sums of functions of $\tilde k_j$ go over to
integrals $A_N^2\sum_jF(\tilde k_j)\to (2/\pi)\int F(\tilde k)d\tilde k$. Using
the {\it linearized} dispersion one obtains for $n>0$
\begin{eqnarray}
\label{clapprox}
\langle u^2_n\rangle_{cl}/a^2&\approx &
\frac{2\eta_{cl}}{\pi}\int_0^\pi\frac{\sin^2(\tilde kn)}{\tilde k^2}
d\tilde k\\ \nonumber
&=& \frac{2\eta_{cl}}{\pi}n\left[ \int_0^\infty \frac{\sin^2u}{u^2}du
 -\int_{n\pi}^\infty \frac{\sin^2u}{u^2}du\right ]\\ \nonumber
&\approx&\eta_{cl}\left(n-\frac{1}{\pi^2}\right)~.
\end{eqnarray}  
The integral from $0$ to $\infty$ can be found in tables.
In the integral from $n\pi$ to $\infty$ one only makes a small
error by replacing $\sin^2u$ by its average value $1/2$
if $n$ is sufficiently large. The additional term
$-\eta_{cl}/\pi^2$ explains the small deviation 
obtained in the numerical evaluation for a finite
chain length.

\noindent The deviation $\langle ( x_n-na)^2\rangle_{cl}  $
 from the lattice position
$na$ increases with $n$ without bound in the infinite chain
limit. Therefore {\it no crystalline order}
exists in the classical description at finite temperatures. 

\noindent The explicit form of  ${\bf C}^{-1}$ also leads
to $\langle (u_n-u_l)^2\rangle_{cl}=\sigma^2_{cl}|n-l|$, used later.\\

 We next consider the quantum fluctuations in the ground state.
In order to compare the fluctuations to the lattice constant
 we introduce the $\hbar$-dependent
quantum mechanical dimensionless ratio
\begin{equation}
\alpha \equiv \frac{\hbar}{mac}=2\frac{\sigma_{qm}^2}{a^2}~.
\end{equation}
If one takes experimental values
of three-dimensional crystals in the typical range\cite{AM} 
$c=4\cdot 10^3$m/sec, $ a=4\cdot 10^{-10}$m and $m=Am_{\rm proton}$
with $A$ the nucleon number
one obtains $\alpha \approx 0.04/A$.  
In order for the harmonic approximation to be valid the
ground state fluctuations of a single particle should be small compared
to $a$ which implies $\alpha \ll 1$, fulfilled for not too light
elements.

\noindent   Only the dimensionless ratio $\eta$ defined as
\begin{equation}
\eta\equiv \frac{\eta_{cl}}{\alpha}=\frac{k_BT}{\hbar\omega_s}
\end{equation}
is allowed to be larger than one.

One obtains the ground state quantum fluctuations 
by putting  $n_B=0$ in  Eq. (\ref{unl}).
Dividing by $a^2$ this yields
\begin{equation}
\langle \hat u_n^2\rangle_0/a^2 
= \frac{\alpha}{N+1/2}\sum_{j=1}^N\frac{\sin^2(\tilde k_j n)}
{2\sin(\tilde k_j/2)}~.
\end{equation}
The numerical evaluation of the sum as a function of $n$ is shown in
 Fig. 2 for $N=950$ and $N=1000$.

\begin{figure}
\label{fighc}
\centering
\epsfig{file=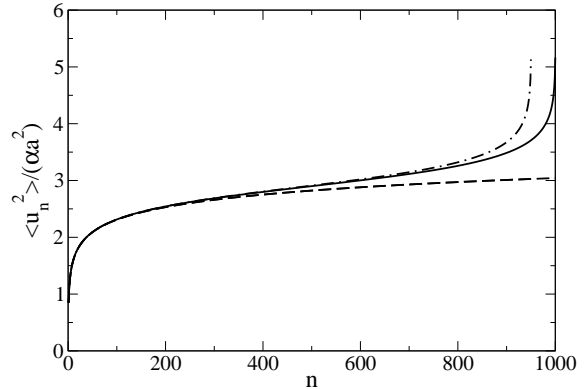,angle=-90,scale=.3}
\vspace{0.2cm}
\caption{Quantum fluctuations  $\langle \hat u_n^2\rangle$ scaled
by $\alpha a^2$ for $N=1000$ (full curve) and  $N=950$ (dashed-dotted curve)
as a function of the distance from the fixed left end of the chain. The
dashed line shows the {\it logarithmic increase}
 $\log n/(2\pi)+\rm{const.}$ discussed in the text.} 
\end{figure}
Again the sum is dominated
by the $0<\tilde k_j \ll 1$ contributions and therefore working
in the numerical evaluation with 
the {\it linearized} dispersion, i.e. $2\sin(\tilde k_j/2) \to \tilde k_j $
again only leads to a small shift ($\approx 0.07$) for
 $ \langle u_n^2\rangle_{cl}/\sigma^2_{qm}$ compared to the
result with the exact dispersion.
The increase with $n$ is much slower
than the linear increase in the classical limit. For $n<N/3$ the
increase can be well approximated by $const.+(1/\pi)\log n$, i.e.
the quantum fluctuations lead to
a {\it logarithmic} increase. This can be seen  analytically in the
infinite chain limit
\begin{eqnarray}
\label{flqm}
  \langle \hat u_n^2\rangle_0/a^2   &\approx &
\frac{\alpha}{\pi}\int_0^\pi\frac{\sin^2(\tilde kn)}{\tilde k}
d\tilde k\\ \nonumber
&=& \frac{\alpha}{\pi}\left [\int_0^b \frac{\sin^2u}{u}du
 +  \int_b^{n\pi} \frac{\sin^2u}{u}du\right ]\\ \nonumber
&\approx& \alpha \left({\rm const.}+\frac{1}{2\pi} \log n\right)~,
\end{eqnarray} 
where in the integral from $b(\approx 1)$ to $n\pi$ we again
 replaced $\sin^2u $ by its average value $1/2$.
In conjunction with the small value of $\alpha$
the slow logarithmic increase of the fluctuations 
 implies that the
long range order due to the {\it quantum fluctuations} is destroyed
 only in {\it very long chains}
with $N\gg e^{\pi/\alpha}$.  A good way to visualize this is to 
consider the ground-state density. Its operator is given by 
\begin{eqnarray}
\label{Gauss2}
\hat \rho(x)&=&\sum_{n=1}^N\delta(x-\hat x_n)\\ \nonumber
&=&\int_{-\infty}^\infty\frac{dk}{2\pi} e^{ikx}
\sum_{n=1}^N e^{-i k\hat x_n}\equiv 
\int_{-\infty}^\infty\frac{dk}{2\pi} e^{ikx}\hat \rho_k~.
\end{eqnarray}
For arbitrary temperatures the expectation value
$ \langle \hat \rho_k \rangle $ can be calculated using 
$\langle e^{\hat A} \rangle=
e^{ \langle \hat A^2 \rangle/2} $.
In order to obtain $   \langle \hat \rho(x)\rangle  $ one has to
perform the Gaussian integral in Eq. (\ref{Gauss2})
\begin{equation}
 \langle \hat \rho(x)\rangle
=\sum_{n=1}^N\frac{1}{\sqrt{2\pi\langle \hat u_n^2\rangle}}
e^{-\frac{1}{2}(x-na)^2/\langle \hat u_n^2\rangle}.
\end{equation}
The ground state density is obtained by using $\langle \hat u_n^2\rangle_0 $.
Whether $\langle \hat \rho(x)\rangle_0$ shows a visible
 lattice periodicity depends
apart from $N$ very sensitively on the coupling constant
$\alpha$.
The ground state density near the
right end of a chain with $90000$ atoms is shown in 
 Fig. 3 for different values of $\alpha$.
\begin{figure}
\label{density}
\centering
\includegraphics[width=6cm,height=5cm]{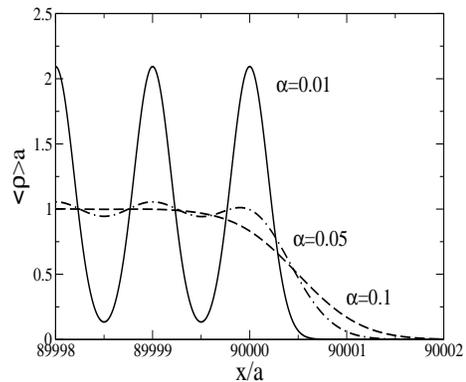}
\vspace{0.5cm}
\caption {Average ground state density of a chain with $90000$
  atoms near the right end of the chain
for different values of $\alpha$. For $\alpha=0.01$ (full curve) 
the lattice structure is still pronounced. It is much weaker for
 $\alpha=0.05$ (dashed-dotted curve) and no longer visible for
  $\alpha=0.1$ (dotted curve).} 
\end{figure}

Now, we address the general finite temperature quantum statistical
mechanics result Eq. (\ref{unl}) for the mean square fluctuations.
 Even at small
temperatures $\eta=k_BT/\hbar \omega_s\ll 1$, the very low energy
modes $\hbar \omega_j\approx \hbar c k_j\ll k_BT$ are highly excited,
i.e. $n_B(\omega_j)\approx k_BT/\hbar \omega_j$. If $n$ is large enough,
there is therefore a contribution
to the $k$-sum dominated by the $1/k_j^2$ behavior
 leading to a linear in $n$ contribution. For small $n$  and  
$\eta=k_BT/\hbar \omega_s\ll 1$ the $T=0$ behaviour dominates. 
The resulting
 crossover from logarithmic to linear increase with $n$ is shown
in Fig. 4 for different values of the (small) temperature. 

In the following subsection the fluctuations $\langle (\hat u_{l+n}-\hat
u_l)^2\rangle$ play an important role which are obtained by replacing 
$(\xi_n^{(j)})^2$ in  Eq. (\ref{unl}) by $(\xi_{n+l}^{(j)}-\xi_l^{(j)} )^2$.
For $l+n$ and $l$ sufficiently far from the boundaries
and $|n|$ sufficiently large 
   $\langle (\hat u_{l+n}-\hat u_l)^2\rangle\approx 
\langle \hat u_n^2\rangle\ $ holds.

\begin{figure}
\label{fighc}
\centering
\epsfig{file=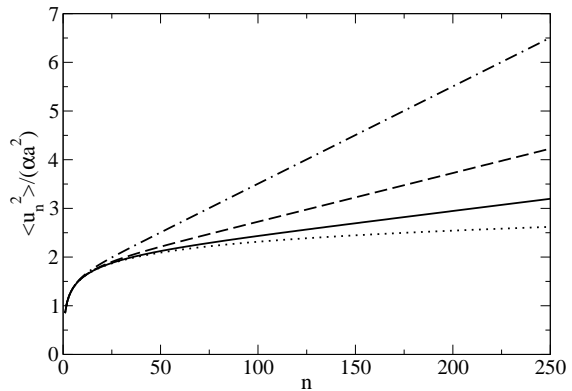,angle=-90,scale=.3}
\vspace{0.2cm}
\caption{Finite temperature quantum statistical mechancis
result for  $\langle \hat u_n^2\rangle$ scaled
by $\alpha a^2$ for $N=1000$ and different small temperatures:
$\eta=0.001$ (dotted line), $\eta=0.0025$ (full line)
  $\eta=0.01$ (dashed line) and $\eta=0.05$ (dashed-dotted line).   }
\end{figure}

\subsection {Experimental implications}

The properties of the chain, like 
the loss of long range order,
can be studied experimentally by the scattering of a test particle with incoming
momentum $\vec p_i$
to a final momentum $\vec p_f$. If the interaction
of the test particle at position $ \vec X$
with the chain atoms is described
by a potential  $V_s=\sum_{i=1}^N v( \vec X-\vec x_n)$, 
 the scattering cross section in Born
approximation involves matrix elements
of the operators $\exp{(i\vec q\cdot \hat{\vec x_n})}$, where 
$\hbar\vec q=\vec p_f-\vec p_i$ 
is the scattering wave vector \cite{AM}. 
 For the chain of atoms aligned on the $x$-axis only $q\equiv q_x$ 
appears in the   
static and dynamic structure factor.
 The {\it static} structure factor $S(q)$ which enters the cross
 section for X-ray scattering
in Born approximation reads  \cite{AM}
\begin{equation}
S_N(q)\equiv \frac{1}{N} \langle \hat \rho_q \hat \rho_{-q}\rangle
=\frac{1}{N}\sum_{n,l=1}^N e^{-iqa(n-l)}\langle e^{-iq
 (\hat u_n-\hat u_l)} \rangle .
\end{equation} 
In order to  evaluate $S(q)$ we use
 $\langle e^{\hat A}\rangle =e^{\langle \hat A^2\rangle/2}$ and obtain
\begin{equation}
\label{staticstructuref}
S_N(q)=\frac{1}{N}\sum_{n,l=1}^N e^{-iqa(n-l)}e^{-q^2
\langle (\hat u_n-\hat u_l)^2 \rangle/2 }.
\end{equation}
In a {\it classical} description the factor containing the
displacements is absent at $T=0$ 
and the structure factor follows performing the geometric series as
\begin{equation}
S_N^{(0)}(q)=\frac{1}{N}\left[ \frac{\sin{(qaN/2)}}{\sin{(qa/2)}}\right ]^2
\end{equation}
For finite $N$ this leads to  {\it Bragg peaks} at $q=2\pi \nu /a\equiv
Q_\nu$ with peak height $N$ and width $1/N$.  
Except for $\nu=0$ (forward scattering) the Bragg peaks   
of the ideal static lattice are weakened by the second exponential
factor on the rhs of Eq. (\ref{staticstructuref})
which involves expectation values discussed in the previous subsection. 
In the following we present numerical results for
long but finite chains as well as analytical results in various
limiting cases.

\noindent The double sum in Eq. (\ref{staticstructuref})
 can be numerically evaluated for finite chains.
Results are shown in Fig. (5) for a chain with $N=1000$ atoms
and $\alpha=0.02$. 
As $S_N(0)=N$ the region near $q=0$ is suppressed.
 The dashed-dotted curve for $\eta_{cl}=0.03$ which is closest to the classical
limit looks similar to the static structure factor of a simple liquid
in three dimensions with 
a (broadened) peak resulting from the typical nearest neighbour
distances \cite{Chandler}.

\noindent This behaviour can be understood
analytically in the strict classical limit $\alpha=0$.
 Then the result for the relative
fluctuation of the chain simplifies to 
  $ \langle (\hat u_n-\hat u_l)^2 \rangle=\eta_{cl} a^2|n-l|$
for {\it all} distances.
\begin{figure}
\label{staticsf}
\centering
\includegraphics[width=6cm,height=5cm]{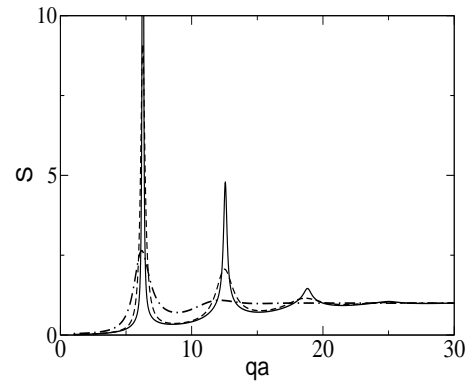}
\vspace{0.5cm}
\caption {Static structure factor for a chain with $N=1000$
atoms  and  $\alpha=\hbar/(mca)=0.02$
for various values of $\eta_{cl}=k_BT/(mc^2)$. The dashed-dotted curve
 for $\eta_{cl}=0.03$ shows ``liquid-like''
behaviour where only a Bragg peak at $q=2\pi/a$ is clearly seen.
 For $\eta_{cl}=0.01$ 
(dashed curve) also the peak at
$q=4\pi/a$ is visible. For the full curve with $\eta_{cl}=0.001$ 
the quantum effects discussed in the text
are important.The forward scattering peak with $S_N(0)=N$ is suppressed.} 
\end{figure}
Therefore
 the double sum in Eq. (\ref{staticstructuref}) 
can be reduced to a single sum using
\begin{equation}
 \frac{1}{N}\sum_{l,n=1}^N G(n-l)=
G(0)+2\sum_{m=1}^{N-1}\left (1-\frac{m}{N}\right) G_e(m),
\end{equation}
where $G_e$ is the {\it even} part of the arbitrary function $G$.
 In the limit $N\to \infty$ the
 geometric series which is convergent for $q\ne0$ leads to \cite{Emery}
\begin{equation}
\label{Scl}
S_\infty^{(cl)}(q)=
\frac{\sinh (\eta_{cl} q^2a^2/2)}{\cosh (\eta_{cl} q^2a^2/2)-\cos(qa)}~.
\end{equation}
As the limits $N\to \infty$ and $q\to 0$ do not commute the forward
scattering peak is lost. 
Near the Bragg positions $Q_\nu$ with nonzero $\nu$, i.e.
$qa=2\pi\nu+\tilde qa$ and  $\tilde qa \ll 1$
one can expand the cosine function in the 
denominator.
 For $\eta_{cl}Q_\nu^2a^2\ll 1$  this leads to well defined 
Lorentzian Bragg peaks of
width $2\pi^2\nu^2\eta_{cl}/a$. 
For finite temperatures this can happen only for small values of
$|\nu|$ and already for $\eta_{cl}=0.01$ the structure factor looks
liquid-like.

\noindent As
the quantum parameter  $\alpha$ is not strictly zero for the results
 in Fig. 5  but
chosen as
 $\alpha=0.02$  the dashed-dotted curve has an additional broadening due to
quantum fluctuations. Reducing the temperature they become more
important for $\eta_{cl}=0.01$
(dotted curve)  and prominent for $\eta_{cl}=0.001$ (full curve).
The peak height at $q=Q_1$ for this parameter value is $\approx 36.7$.

\noindent For small values of $\alpha$ and $T=0$ one has to work with
 very long chains to 
see the {\it divergent power law shape} of the low index Bragg peaks
 which emerges in the infinite chain limit.
In order to determine the divergent contribution 
one can approximate the sum by using 
the long distance logarithmic dependence of
 $\langle (\hat u_{l+n}-\hat u_l)^2 \rangle$ for {\it all} distances 
$|n|>n_0$, where $n_0>1$ can be chosen arbitrarily.
 Using $e^{-\beta \log{n}}=1/n^\beta$ one obtains
for the possibly divergent contribution (indicated by ``$div$'' on top
of the equal sign)
\begin{equation}
\label{Sdiv}
S_\infty(q) \stackrel {div}{=}2\lim_{N\to \infty}
\sum_{n=n_0}^N \cos( qan)\left (1-\frac{n}{N}\right ) \frac{1}{n^{\beta(q)}}  ~ ,
\end{equation} 
where
\begin{equation}
\label{beta}
\beta(q)\equiv \frac{\alpha}{2\pi} (aq)^2.
\end{equation} 
At the Bragg positions  $q=Q_\nu$, the sum is finite 
 for $\beta(Q_\nu)=2\pi\alpha \nu^2>1$ and diverges
for $2\pi\alpha \nu^2\le 1$. As we assume  $\alpha \ll 1$
the sum  $S_\infty(Q_\nu)$ diverges 
 for a few low index Bragg peaks,
but is finite for sufficiently large $|\nu|$.
In order to determine the divergence with $N$ at the low index Bragg
 positions $Q_\nu$ with $\nu\ne 0$ the sum
in Eq.(\ref{Sdiv}) can be converted to an integral which can easily
 be performed as the cosine factor
 equals $1$. Both terms have the same divergent behaviour, only 
the prefactors differ. For $0< \beta(Q_\nu) < 1$ one obtains 
\begin{equation}
T=0:~~~~~~~~~~S_N(Q_\nu) \sim N^{1-2\pi \alpha \nu^2}~.
\end{equation} 
For   $\beta(Q_\nu) < 1$
the divergent peak shape for $q=Q_\nu+\tilde q$
in the small $\tilde qa$ limit
is also determined by the large $n$ behaviour of the sum.
 In order to read-off the divergent
contribution to $ S_\infty(Q_\nu+\tilde q) $
 for $\nu\ne 0$ one can again replace the sum by an integral.
 This yields after the substitution $u=|\tilde q|an$
\begin{equation} 
\label{pl1}
 S_\infty(Q_\nu+\tilde q)   \stackrel {div}{=}   2
(|\tilde q|a)^{\beta(q)-1}\int_{\tilde q an_0}^\infty
\frac{\cos u}{u^{ \beta(q)}}du~.
\end{equation} 
 Because of the cosine factor
 the upper limit of the integral poses no problem as $\beta(q)>0$. 
For   $\beta(q)<1$  the integral is also non-divergent at the lower 
limit for $\tilde q \to 0$.
In this limit $\beta(q)$ in the prefactor can be replaced by  $\beta(Q_\nu)$.
This leads to the
 {\it power law divergence} of the static structure factor near the
 Bragg peaks with  $\beta(Q_\nu)<1$
\begin{equation}
\label{pl2}
  S_\infty(Q_\nu+\tilde q)     \sim (|\tilde q|a)^{\beta(Q_\nu)-1}
=(|\tilde q|a)^{2\pi\alpha \nu^2-1}.
\end{equation}
\noindent  As we have only
focussed on the divergent part of $S_\infty$ in Eq. (\ref{pl1}) the
power law divergence
 of Eq. (\ref{pl2}) does {\it not} imply that
$S_\infty$ vanishes in a power law fashion in $|\tilde q|$ at the
 Bragg positions with  $\beta(Q_\nu) >1$.\\

As a second experimental probe we shortly discuss 
 M\"ossbauer spectroscopy which is a
 {\it local} probe \cite{Lipkin}. We assume that the nuclei of 
the atoms in the chain have an excited state which can decay to the
 ground state by emitting a gamma ray. In this process
the momentum $-\hbar \vec q $ is transferred to the atom which emits the gamma 
quantum with wavevector $\vec q$. 
 The energy of the emitted gamma
 quantum depends on the recoil energy transmitted to the crystal. 
 There is a finite probability that the emitted gamma
ray is absorbed by another nucleus of the same type. This
{\it recoilless} emission is called  ``M\"ossbauer effect''.
If the harmonic crystal is in an eigenstate $|E_n\rangle $ initially
 (with $n$  a multi-index), the state after the gamma emission
from the atom with the position $\vec x_l$ is given by
 $\exp{(-i\vec q\cdot \hat {\vec x_l})}|E_n\rangle $.
The energy spectrum of the emitted gamma ray relative to its energy
for  a fixed nucleus is obtained by
expanding this state into eigenstates of the harmonic crystal and
averaging over the canonical probability
for the initial state  $|E_n\rangle $,
\begin{equation}
P_l(\epsilon)= 
\sum_{n,m}\frac{e^{-\beta E_n}}{Z}
|\langle E_m|e^{-i\vec q\cdot \hat {\vec x_l}}|E_n\rangle|^2
\delta(\epsilon +(E_m-E_n)).
 \end{equation}
Using the Fourier representation of the delta function
it is straightforward to calculate this function for the
harmonic chain with the help of the operator identities presented in the
previous sections. 

Here we only address the question how
at $T=0$ the
 quantum fluctuations suppress recoilless emission.
For the one-dimensional chain only $q\equiv q_x$ enters and
the probability not to excite phonons is given by
\begin{equation}
p_l^{(0)}=|\langle E_0|e^{-iq  \hat u_l}|E_0\rangle|^2
=e^{-q^2\langle \hat u_l^2\rangle_0}~.
 \end{equation}
Using the infinite chain result Eq. (\ref{flqm}) one finds that
 the probability for recoilless emission decreases as a power
 law with the distance from the fixed left end of the chain.
\begin{equation}
p_l^{(0)}\sim l^{-\beta(q)}~,
 \end{equation}
with $\beta(q)$ defined in Eq. (\ref{beta}).\\

As a third experimental probe neutron scattering could be discussed
with the help of the dynamic structure factor.\cite{AM,Mikeska}.

\section{Outlook}

We have shown how classical thermal fluctuations and quantum
fluctuations at $T=0$ destroy the long range order in an infinite
chain in a very different way. Both results
are caused by the low energy
phonon modes with frequency $\omega(k) \approx ck$. In the classical
high temperature limit the {\it linear} divergence of 
$\langle \hat u_n^2 \rangle $ with $n$ is due to the $1/k^2$
divergence of the integrand in Eq. (\ref{clapprox}). The factor 
$\sin^2(kn)$ leads to a small $k$ cutoff  $k_{\rm min}\sim 1/n$ 
for the $k$-integration.
At $T=0$ the quantum fluctuations follow from the  $1/k$ divergence
of the integrand in  Eq. (\ref{flqm}), the cutoff being the same.
This leads to the slow {\it logarithmic} divergence of 
$\langle \hat u_n^2 \rangle $ with $n$, i.e. crystalline long
range order is destroyed even at zero temperature.

The findings for the infinite
one-dimensional chain can  be generalized to higher dimensional harmonic
lattices by putting the atoms e.g. on a square lattice or a simple
cubic lattice. The low frequency modes are again sound waves\cite{AM} with
$\omega(\vec k)\approx c|\vec k|\equiv ck$. The mean square deviations
can be calculated as for the one-dimensional chain in Eq. (\ref{unl}).
The possibly diverging part again comes from the small $k$
contribution with the $dk$ in the infinite chain limit replaced
by  the $d$-dimensional 
 integration measure proportional to $k^{d-1}dk$.
 Multiplying with  $1/k^2$ for the 
thermal fluctuations yields for the possibly diverging contribution
$\sim \int_{k_{min}}k^{d-3}dk$ and multiplying with  $1/k$ 
for the quantum fluctuations at $T=0$ to
 $\sim \int_{k_{min}}k^{d-2}dk$. In the three-dimensional case
even the thermal fluctuations lead to a finite result in the limit
$k_{min}\to 0$ i.e. long range crystalline order exists as 
noted by Peierls \cite{Peierls}.
 For $d=2$
the quantum fluctuations are finite in this limit but a {\it
  logarithmic} divergence is produced by the thermal fluctuations.
 The divergence due
to quantum fluctuations in $d$ dimensions is of the same type
as from the (classical) thermal fluctuations in $d+1$ dimensions,
where in our example $d=1$. 

Extensions of this finding for harmonic lattices can be found in the
literature on quantum phase transitions \cite{Sachdev}. This are phase
 transitions not as a function of tempertature but as a function of 
a system parameter at zero temperature. The harmonic chain with the
additional harmonic potentials of strenght $\lambda_ {loc}$ centered at the
lattice positions discussed following Eq. (\ref{lowenergy}) has a quantum
phase transition at  $\lambda_ {loc}=0$. For all positive values of 
 $\lambda_ {loc}$ the fluctuations around the lattice positions stay finite
and well defined Bragg peaks exist. For  $\lambda_ {loc}=0$ the 
zero temperature quantum fluctuctions diverge logarithmically and
no crystalline order exists. This ``one-sided'' quantum phase transition
 is of a very special type as only values   $\lambda_ {loc}\ge 0$ 
correspond to stable systems.

\section{Acknowledgements}

The author would like to thank G. Hegerfeldt, V. Meden, H.J. Mikeska
and D. Vollhardt for a critical reading
of the manuscript.

\end{document}